\documentclass[onecolumn,letterpaper,amsmath,amssymb,floatfix,aps,preprintnumbers,superscriptaddress]{revtex4}

\usepackage{verbatim}
\usepackage{graphicx}
\usepackage{bm} 
\usepackage{dcolumn}  
\usepackage{epsfig}
\usepackage{color}

\begin{document}

\title{Aspects of One-Dimensional Coulomb Gases}

\author{Ronald~R. \surname{Horgan}}
\affiliation{Department~of~Applied~Mathematics~and~Theoretical~Physics,
University~of~Cambridge, Centre~for~Mathematical~Sciences, Cambridge~CB3~0WA, United~Kingdom}
\author{David S. \surname{Dean}}
\affiliation{Universit\'e de  Bordeaux and CNRS, Laboratoire Ondes et Mati\`ere d'Aquitaine (LOMA), 
UMR 5798, F-33400 Talence, France}
\author{Vincent \surname{D\'emery}}
\affiliation{Institut Jean Le Rond d'Alembert, CNRS and UPMC Univ Paris 6, UMR 7190, F-75005 Paris, France}
\author{Thomas~C. \surname{Hammant}}
\affiliation{Department~of~Applied~Mathematics~and~Theoretical~Physics,
University~of~Cambridge, Centre~for~Mathematical~Sciences, Cambridge~CB3~0WA, United~Kingdom}
\author{Ali \surname{Naji}}
\affiliation{School of Physics, Institute for Research in
Fundamental Sciences (IPM), Tehran 19395-5531, Iran }
\affiliation{Department~of~Applied~Mathematics~and~Theoretical~Physics,
University~of~Cambridge, Centre~for~Mathematical~Sciences, Cambridge~CB3~0WA, United~Kingdom}
\author{Rudolf \surname{Podgornik}}
\affiliation{Department of Physics, Faculty of Mathematics and Physics, University of Ljubljana, SI-1000 Ljubljana, Slovenia}

\begin{abstract} 
In this short review, we discuss recent advances in exact solutions of models based on a one-dimensional (1D) Coulomb gas by means of field-theoretic functional integral methods. 
The exact solutions can be used to assess the accuracy of various approximations such as the weak coupling Poisson-Boltzmann theory as well as the  strong coupling theory
of Coulomb gases. We consider three different 1D models: the Coulomb fluid configuration in the case of the soap film model consisting of positively and negatively charged particles between adsorbing boundaries, counterions between two charged surfaces, and  an ionic liquid lattice capacitor with positively and negatively charged  particles on a lattice between one positive and one negative bounding surface. 
\end{abstract}

\maketitle

\section{Introduction}

Field-theoretic functional integral methods can be used to study exact solutions of models based on a one-dimensional (1D) Coulomb gas with charged boundaries. In 1D, exactly solvable Coulomb gas models can be then used as a testbed for assessing the accuracy of various approximations: the weak coupling expansion, Poisson-Boltzmann/mean-field equations, and the strong coupling expansion \cite{netz:2005}. We review these approximations in the context of three 1D Coulomb gas systems and remark on whether or not they fail to predict important effects present in the exact solution. 

Some physical properties of the 1D system can be applicable at least qualitatively for dimensions $d > 1$ and can help us to understand whether pertaining approximation methods are reliable or not. In particular, our analysis gives insight into  systems such as an array of charged smectic layers or lipid multilayers, and ionic liquids near charged interfaces, treated as effectively 1D systems. An important aspect of these endeavours is that we can test and develop the analysis and especially numerical methods that can then be tentatively applied also for $d > 1$.

\begin{figure}[h]
\centerline{\psfig{file=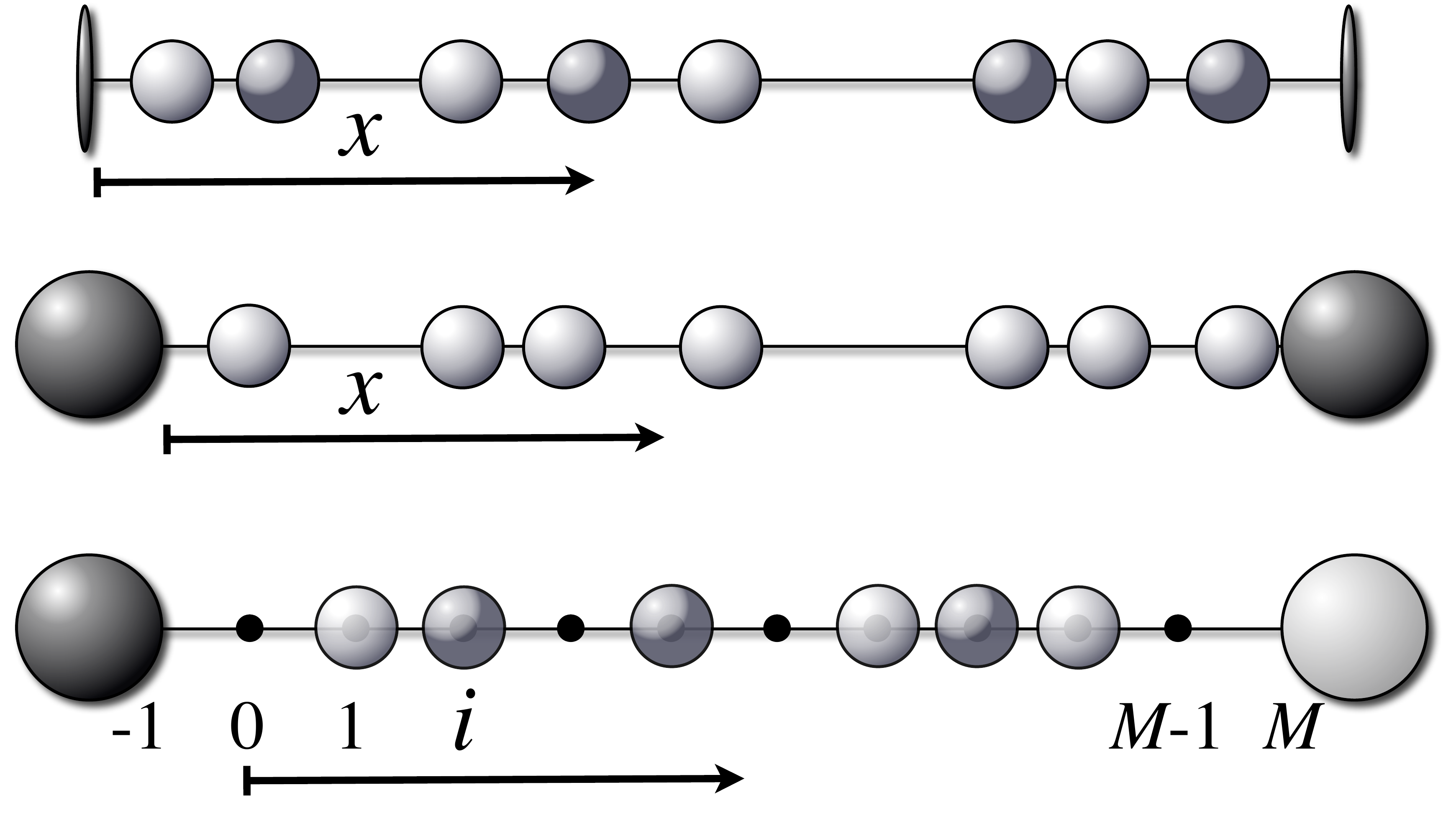,width=6cm}}
\caption{Top: The 1D Coulomb fluid configuration in the case of the soap film model (positively and negatively charged particles and adsorbing surfaces), middle: counterions between charged surfaces (positively charged particles and oppositely charged bounding surfaces) and bottom: an ionic liquid lattice capacitor (positively and negatively charged  particles on a lattice with one positive and one negative bounding surface). Positively (negatively) charged particles are shown schematically as light (dark) gray spheres. }
\label{scheme}
\end{figure}

\section{\label{th_meth} Theoretical Methods}

The method of functional integrals applied to Coulomb gas systems has been developed over many years 
\cite{EdwLen:1962,PodZek:1988,Dea_ea:1998,NetOrl:1999}. In any dimension this approach allows for both strong and  weak coupling to be studied explicitly, but specifically in 1D the functional integral representation can be applied using a variety of methods to obtain exact solutions to a number of models which are generally characterized by a Coulomb gas of ions of possibly non-zero size confined between boundaries with properties that allow their potential or charge to be determined either dynamically or as an external field condition. Three varieties of a 1D Coulomb gas model discussed below are presented in Fig. \ref{scheme}.

The functional integral representation of the Coulomb gas partition function allows us to formulate two effective solution techniques.  The {\em Schr\"odinger kernel technique}  is applicable in all dimensions and has 
been used to analyze a number of models \cite{DeaHor:2005}. In 1D it
corresponds to solving the Schr\"odinger equation \cite{EdwLen:1962} which is in principle exact. In $d>1$ the Schr\"odinger kernel field theoretic representation of the partition function is derived, often using a Hubbard-Stratonovich transformation, and is
analyzed by perturbative and graphical methods. For $d>1$ this approach does require that a preferred co-ordinate 
can be designated as the Euclidean time and so the approach is limited to symmetrically layered systems \cite{DeaHor:2005}. The  {\em transfer matrix and Fourier methods technique} is an alternative to the Schr\"odinger 
kernel approach. Though it is more general, it is only practical in 1D. Its implementation exploits periodicity in the 
(imaginary) electrostatic potential $\phi$ which also restricts its general applicability. Examples of this technique in 1D are the counterion gas and lattice ionic liquids.


The actual formulation of the functional integral method relies on the action for the full QED of a general system that is then reduced to the electrostatic action proper. The relevant {\em electrostatic Lagrangian} is then
\begin{equation}
{\cal L}(\psi) = \frac{1}{2}\varepsilon\int d{\bf x} (\nabla\psi({\bf x}))^2 - e\sum_iq_i\psi({\bf x}_i)
- \int d{\bf x}\rho_e({\bf x})\psi({\bf x}).
\nonumber
\end{equation}
Here $ q_i$ is the charge of the $i$-th ion  at position $ {\bf x}_i$ and $\rho_e({\bf x})$ is the external charge
distribution. The partition function is obtained by tracing the Boltzmann weight of the above Lagrangian over the electrostatic field $[\psi]$.
Tracing furthermore over ion positions, changing the axis of functional integration $ \psi = i\phi~$ and introducing fugacity $\mu_-=\mu_+\equiv \mu$ by the Gibbs technique, the partition function for monovalent ions (with $q_i=\pm1$) assumes the form
\begin{eqnarray}
 {\cal Z}& =& \int\;d[\phi]\exp(S(\phi))\;, 
 \end{eqnarray}
with the  ``field action"
\begin{eqnarray}
 S(\phi)& =& -\frac{\beta}{2}\varepsilon \int d{\bf x} (\nabla \phi({\bf x}))^2 +  2\mu \int d{\bf x}\;\cos(e\beta\phi({\bf x}))\;, 
\end{eqnarray}
where $\beta=1/(k_BT)$. The charge density operator is then given by
\begin{equation}
\rho =\mu \frac{d}{d\mu}\log {\cal Z}(\mu)~~\Rightarrow~~\rho = 2\mu\langle\cos(e\beta\phi)\rangle,
\label{cdensity}
\end{equation}
where $\langle \cdots \rangle$ stands for the $\phi$ average.

\section{\label{soap_film}Bilayer soap film in ionic solution}
Because hydrophobic heads of the surfactant molecules  preferentially migrate to the surfaces charging them up dynamically, the configuration of the bilayer soap film consists of two planar (surfactant) surfaces separated by a distance $L$ confining a solution of a symmetric electrolyte. 

We calculate the surface charge, the density profile of electrolyte near the interfaces, and the disjoining pressure $P$ as a function of the thickness $L$ of the soap film, defined as
\[
 P = P_\mathrm{film}-P_\mathrm{bulk}  = -\frac{1}{\beta}\left(\frac{\partial J_\mathrm{film}}{\partial L} -
                   \frac{\partial J_\mathrm{bulk}}{\partial L}\right).
\]
{\em i.e.} the difference between the film and bulk pressures. Here $ J$ is the grand-canonical partition function/unit area. An important phenomenon to predict is the first-order collapse transition of the film to a Newton black film expected as the electrostatic coupling in the film is increased. 


We model this system by a Coulomb gas confined to $z \in [0,L]$, schematically presented in Fig. \ref{scheme} (top), with potentials on the boundaries that account for the hydrophillic nature of the head group of the surfactant molecule. The Debye length is given by $l_D=\sqrt{{\varepsilon k_BT}/{2\rho e^2}},$ and the  Bjerrum length in 1D by $l_B={2k_BT\varepsilon}/{e^2}$. Perturbation theory is an expansion in the coupling parameter $g = l_D/l_B$. 

We use the partition function described earlier but now includes surface free energy $f(\phi)$ to model the surface potentials, which are attractive for the negatively charged hydrophillic surfactant head groups whose surface density is denoted by $\rho_-(\phi)$:
\[
f(\phi) = e^{\textstyle \lambda\rho_-(\phi)},
\]
where $\lambda$ controls the potential strength. To simplify the notation we scale the variables: $\phi \to e\beta\phi$,~~${\bf x} \to {\bf x}\, l_B$. The charge density operators for $\pm$ charges are then given by the Boltzmann weights $\rho_\pm(\phi) = e^{\textstyle \pm i\phi}$. The 1D partition function then becomes
\[
{\cal Z} = \frac{1}{2\pi}\int_0^{2\pi}d\phi_0d\phi_L\, f(\phi_0)K(\phi_0,\phi_L;L)f(\phi_L)\;,
\]
where 
\[
K(\phi_0,\phi_x;x)~=~\int{\cal D}\phi(x)\, \exp{\int_0^x dx'\,{\cal L}(\phi(x'))}
\]
is the Schr\"odinger kernel for evolution in the ``Euclidean time'' $x$:
\[
\Psi(\phi,x) = \int d\phi'\,K(\phi,\phi';x)f(\phi').
\]
It satisfies the Schr\"odinger (Feynman-Kac) equation
\[
H\Psi = \frac{2\varepsilon}{\beta e^2}\frac{\partial}{\partial x}\Psi\;,~~~~~~H = \frac{\partial^2}{\partial\phi^2}~+~
\frac{Z(g)}{2g^2}\cos(\phi),
\]
with $Z(g) = 1/\langle \cos(\phi) \rangle$. The above equation is also known as the {\em Mathieu equation}, and the harmonic term gives the Debye length in units of $l_B$.  $Z(g) = 2\mu/\rho$ is the renormalization that relates the fugacity to the observable charge density and is given by Eq. (\ref{cdensity}). We now consider the solution in various limiting regimes.

\subsection{Large $L$: bulk pressure}

{\bf Strong coupling} (SC) $g \to \infty$: 
The Mathieu ground state dominates in this regime and so we can use the Schr\"odinger perturbation theory for the ground-state energy of $H$. The result, derived originally in \cite{EdwLen:1962}, is
\[
P_\mathrm{bulk} = \frac{1}{2}\rho k_BT\left[1+\frac{7}{32}\frac{1}{g^2}-\frac{23}{4608}\frac{1}{g^4} -
\frac{4897}{7826432}\frac{1}{g^6}+\ldots\right].
\]
The leading term is the free gas term but for density $\rho/2$, which therefore signals the onset of the dimerization process, {\em i.e.}, the Bjerrum pair formation of positive and negative mobile charges.

\noindent{\bf Weak coupling}  (WC) $g \to 0$: Feynman perturbation theory is applicable in this case and so we use the Feynman diagram expansion to find
\[
P_\mathrm{bulk} = \rho k_BT\left(1 - \frac{1}{2}g + \frac{1}{128}g^3 + \ldots\right).
\]
The leading term is the free gas term and the second-order term is the familiar Debye-H\"uckel result in its 1D variant. Note that there is no $O(g^2)$ term; this is cancelled by the counter term in $Z(g)$. 

The strong and weak coupling dependencies of the bulk pressure $P_\mathrm{bulk}$ on $g$ compare well with the exact solution of the problem. Both approximations are accurate across a wide range of $g$ in their regime of validity. More details can be found in \cite{Dea_ea:1998}.


\subsection{Finite $L$: exact methods}

For finite $L$ we expand the kernel $K(\phi_0,\phi_L;L)$ over periodic eigenfunctions of the Mathieu equation. We can then use a numerical approach for eigenfunctions/eigenenergies which will give an exact solution for all $L$. This method is described fully in \cite{Dea_ea:1998} and we do not delve into details here. It gives the same answers as the Fourier approach that we describe below.

The Fourier method for obtaining an exact solution to problems in 1D is more general than the Schr\"odinger approach since it works also when the Hamiltonian is not hermitian, which is the case for the counterion gas considered in the next section. It also forms the basis for the transfer matrix method. The theory is periodic under $~\phi \to \phi+2\pi$ and we can define
\[
\Psi(\phi,x) = \int K(\phi,\phi';x)f(\phi')\;d\phi' = e^{\textstyle \;x/2g^2}\sum_{n=-\infty}^{n=\infty} b_n(x)e^{\textstyle in\phi},
\]
where the coefficients $b_n(x)$ obey the evolution equation
\[
\frac{db_n}{dx} = -n^2b_n + \frac{1}{4g^2}(b_{n+1} + b_{n-1} - 2b_n).
\]
This is the Fourier version of the Schr\"odinger equation but can be derived generally from the convolution property of the Schr\"odinger kernel. 
The partition function can then be obtained from
\[
{\cal Z}(T,L) = \int_0^{2\pi} f(\phi)\Psi(\phi,L)\;d\phi~=~e^{\textstyle \;L/2g^2}\sum_{n=0}^{n=\infty} \frac{\lambda^n}{n!}b_n(L),
\]
The exact solution for the disjoining pressure as a function of the separation $L$ for different values of the surface potential strength parameter $\lambda$ clearly predicts a collapse transition to a Newton black film that can not be accounted for by the mean-field theory, which we address next. 


\subsection{Classical or mean-field (MF) theory} 

Standard variational methods applied to the expression for the partition function gives the classical MF equation: the Poisson-Boltzmann (PB) equation for $\phi_{cl}(x)$ as the saddle point equation of the corresponding field theory. In this case the disjoining pressure $P$ is given by the value of ion density at the midpoint $x=L/2$ between the bounding surfaces. 
The MF theory predicts that universally $P > 0$, 
contrary to our exact result and also to experiment; it does not predict any collapse transition, which is thus obviously a consequence of the non-MF correlation effects and is intrinsically a  fluctuation phenomenon. 

\section{\label{counterions}Counterions between Charged Surfaces}

The 1D model here is a Coulomb gas of counterions confined between two oppositely charged surfaces; the  system is overall neutral.  We compare exact results with strong and weak coupling calculations, which are the same as in a 3D system. More details can be found in \cite{Dea_ea:2009}.  

The system is shown in Fig. \ref{scheme} (middle) and consists of $N$ counterions, each of valency $q$, with surface charges $\sigma_1$ and $\sigma_2$, respectively.  We define $\zeta = \sigma_2/\sigma_1$, with $ -1 < \zeta < 1$, and define $ \alpha = 1/(1+\zeta)$. The 1D Bjerrum length is $l_B= {2k_BT\varepsilon}/{e^2}$, and the Gouy-Chapman length is $ \mu \equiv \mu_1 = l_Be/q|\sigma_1|$, where we have chosen $\sigma_1$ to be non-zero and have $\mu_2 = \mu/|\zeta|$.  The electrostatic coupling constant, $g$, is then given by
\[
 g \equiv \frac{q^2\mu}{l_B} = \frac{1+\zeta}{N}\;,
\]
where  $N \to \infty$ corresponds to the MF/PB theory and $N \to 1$ to the SC theory. The partition function is derived as
\[
{\cal Z}_N = \frac{1}{2\pi}\int_0^{2\pi}\!\!\!\!\!\!d\phi(0)\int_{-\infty}^{\infty}\!\!\!\!\!\!d\phi(L)\;
e^{\textstyle i\frac{\sigma_1}{qe}\phi(0)}\left(\int_{\phi(0)}^{\phi(L)}\!\!\!\!\!\!d\phi\; e^{\textstyle -S(\phi)}\right)\;e^{\textstyle i\frac{\sigma_2}{qe}\phi(L)},
\]
with
\[
S(\phi) = \int_0^L dx\; \left[\frac{1}{2q^2e^2\beta}\left(\frac{d\phi(x)}{dx}\right)^2 - e^{\textstyle i\phi(x)}\right]\;.
\]
The integral over $ d\phi(0)$ ensures charge neutrality: $ Nqe + \sigma_1+\sigma_2 = 0$.  The corresponding Hamiltonian (Feynman-Kac) and the partition function are then
\[
 H = -\frac{q^2e^2\beta^2}{2}\frac{d^2}{d\phi^2} - e^{\textstyle i\phi}, ~~~~ 
{\cal Z}_N = \int_0^{2\pi}d\phi\;e^{\textstyle i\frac{\sigma_1}{qe}\phi}\;\left[ e^{-LH}\right]\;e^{\textstyle i\frac{\sigma_2}{qe}\phi}.
\]
\subsection{\label{subsec:exact_p}Exact results}
In this system $ H$ is not hermitian because the counterions are, by definition, of one charge only. We therefore analyze the model  using the Fourier method. We exploit periodicity in  $ H$ of $ ~\phi \to \phi+2\pi~$ in order to write
\[
f(\phi;x) = e^{\textstyle -xH}f(\phi;0)~~~~\mbox{with}~~~~f(\phi;x) = \sum\;b(n,x)e^{\textstyle in\phi}\;.
\] 
By introducing  $ \sigma_1 =-M_1q e$, $ \sigma_2 =-M_2q e$, $M_1 = \mbox{Int}(\alpha N)$, $M_2 = N-M_1-1$, $\eta_1=\alpha N-M_1$, and $\eta_2=1-\eta_1$ and 
$\alpha=1/(1+\zeta)$, 
we can derive the Fourier evolution equation from surface $ 2$ to surface $ 1$ in the form
\begin{eqnarray}
\frac{db(n,M_2,L)}{dL}&=&-\frac{(n-\eta_2)^2}{2}\beta q^2e^2b(n,M_2,L) + b(n-1,M_2,L), ~~~~~~
\end{eqnarray}
with $b(n,M_2,0) = \delta_{n,-M_2}$. 
This Fourier evolution equation can be integrated numerically and the corresponding partition function 
\[
Z(\sigma_1,\zeta,N,L) =  b(M_1+1,M_2,L),
\]
and disjoining pressure 
\[
P(L) = -\frac{(M_1+1-\eta_2)^2}{2}q^2e^2~+~\frac{b(M_1,M_2,L)}{Z(\sigma_1,\zeta,N,L)} 
\]
can be evaluated exactly. Since the second term in the above equation can be seen to be just the counterion density at the boundary of the system, the above form of the pressure is thus a clear example of the contact value theorem; it  connects the pressure with the value of the particle density at the confining wall of the system.


\subsection{\label{subsec:wc_p}Weak coupling}

We consider the WC expansion $ g \to 0$ which is equivalent in the lowest order to the MF/PB theory. In the $d>1$ case, the MF theory treats the potential field $\phi({\bf x})$ as constant in the directions transverse to the normal to the bounding interfaces, and so the results are independent of the dimensionality. 

The leading contribution arises from the saddle-point configuration $ \phi_0(x) = i\psi_0(x)$ with $\psi_0$ real. The PB equation and the boundary conditions have the form
\[
\frac{d^2\psi_0(x)}{dx^2} = -q^2e^2\beta\,e^{\textstyle -\psi_0(x)}\;,
\]
with
\[
\frac{d\psi_0}{dx}\bigg|_{0}=-\sigma_{1}\beta qe,~~~~~~ \frac{d\psi_0}{dx}\bigg|_{L}=\sigma_{2}\beta qe. 
\]
The leading PB contribution to the disjoining pressure, $P$ , is then expressed as 
\[
\beta P = -\frac{1}{2q^2e^2\beta}\left(\frac{d\psi_0}{dx}\right)^2 + \rho_0(x),
\]
where $\rho_0(x)$ is the  density of counterions between the boundaries, given by the standard Boltzmann form $\rho_0(x) = C e^{\textstyle-\psi_0(x)}$, where $C$ is a normalization constant. 
This furthermore implies that the MF/PB disjoining pressure 
$P$ is obtained as follows: When the pressure is repulsive ($P>0$), we have $P = {\mu^2\sigma_1^2 \Gamma^2}/{2}$, where $\Gamma$ satisfies 
\[
\tan(\Gamma L) = \frac{\Gamma(1+\zeta)\mu}{\Gamma^2\mu^2-\zeta},
\]
and when the pressure is attractive ($P<0$), which may be the case within the MF/PB theory only for $\zeta<0$, we have $P = -{\mu^2\sigma_1^2 \Gamma^2}/{2}$, where $\Gamma$ is now given as a solution of
\[
~~~~\coth(\Gamma L) = -\frac{\zeta+\mu^2\Gamma^2}{\mu \Gamma(1+\zeta)}.
\]

\subsection{\label{subsec:sc_p}Strong coupling}
The strong coupling limit is formally identical to the one-particle limit \cite{netz:2005}. In the present case it is easily evaluated from the partition function in the case of a single counterion in the system. The partition function in an explicit one-particle form leads to the disjoining pressure 
\[
P(L) = \frac{\sigma_1^2}{2}\left(-\frac{1}{2}(1+\zeta^2) + \frac{1}{2}(1-\zeta^2)\coth\left[(1-\zeta)\frac{L}{2\mu}\right]\right).
\]
The range of validity of this limiting expression is of course defined by the number of counterions in the system. As this number decreases towards one, $N \rightarrow 1$, the above expression for the disjoining pressure becomes exact.

\begin{figure*}[t!]\begin{center}
\begin{minipage}[b]{0.45\textwidth}\begin{center}
\includegraphics[width=\textwidth]{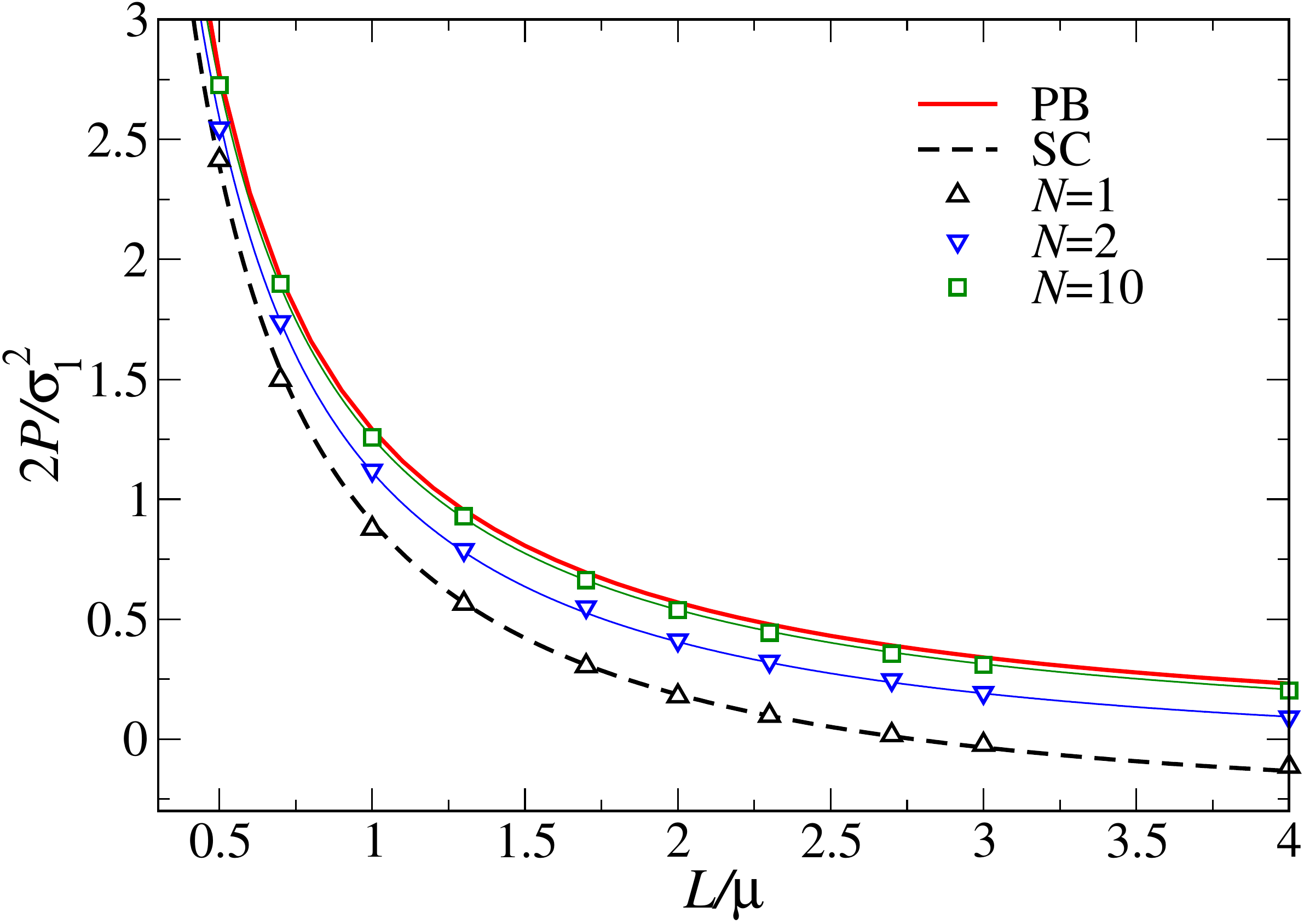} 
\end{center}\end{minipage}
\begin{minipage}[b]{0.45\textwidth}\begin{center}
\includegraphics[width=\textwidth]{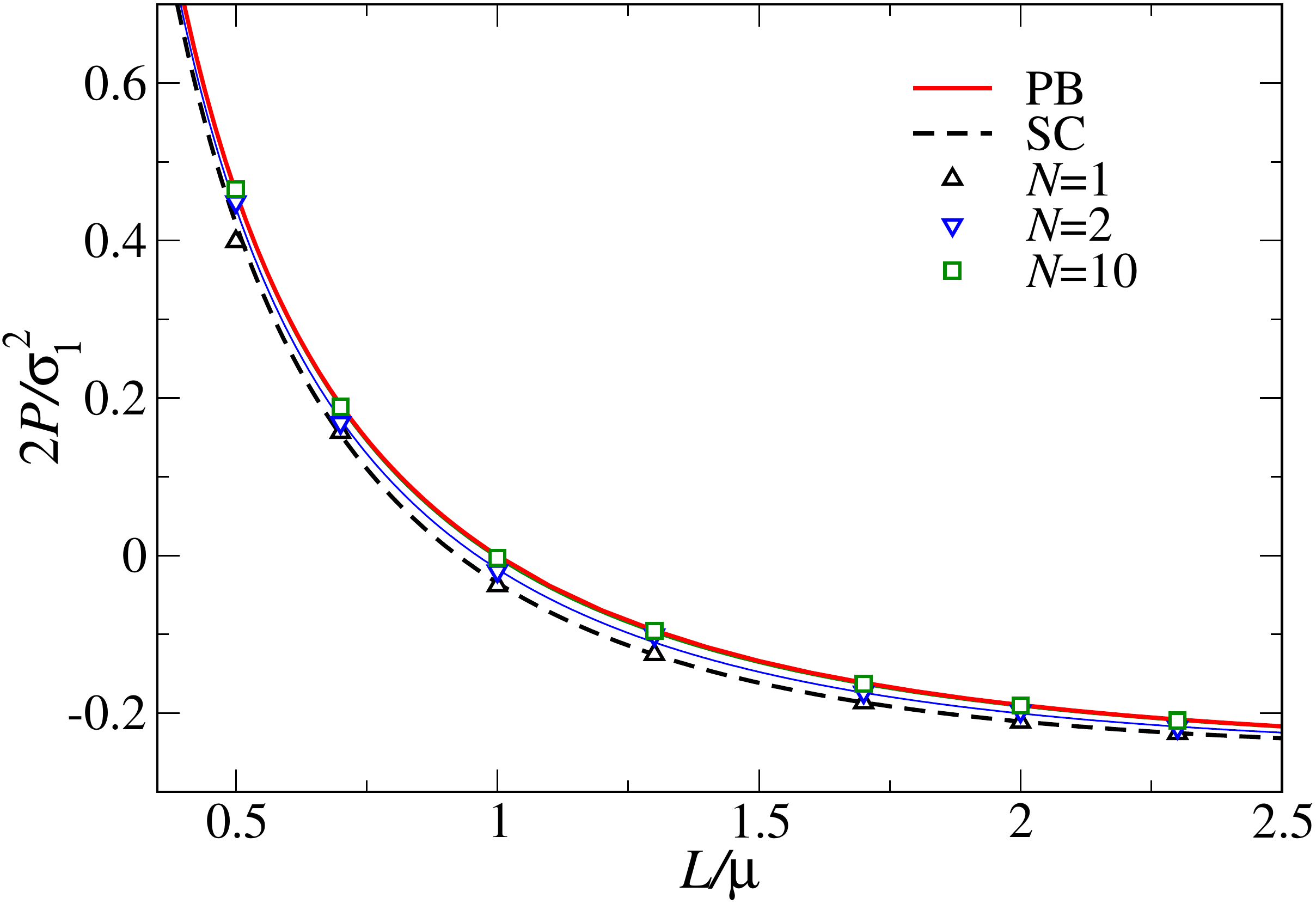} 
\end{center}\end{minipage}
\caption{Rescaled disjoining pressure, $2P/\sigma_1^2$, for a 1D counterion gas between charged surfaces as function of the rescaled intersurface separation $L/\mu$. Thick (red) solid lines represent the PB result (Section \ref{subsec:wc_p}), dashed lines are the SC results (Section \ref{subsec:sc_p}) and thin solid lines are the exact results (Section \ref{subsec:exact_p}) compared with  MC simulations data (symbols) at different numbers of counterions $N$ and for $\zeta = 0.5$  (left pane) and $\zeta = -0.5$  (right pane) \cite{Dea_ea:2009}.}
\label{fig:counter}
\end{center}\end{figure*}

\subsection{Comparison}
Both the weak  and strong coupling  approximations are independent of dimension $d$ and the comparison with the exact results can test their validity.  For symmetric surface charges ($\zeta = 1$) the PB/MF pressure is positive (repulsive) for all intersurface separations, whereas the SC expansion and the exact result for $N=1$ predict attraction at large separations; this distinction holds for $0 < \zeta \le 1$.  For the asymmetric configuration with $\zeta < 0$, there is little difference between the different approaches; on trivial grounds there is attraction for large separations but there is repulsion for sufficiently small separations, see Fig. \ref{fig:counter}, where a comparison is made with Monte-Carlo (MC) simulations  
at different numbers of counterions $N$ \cite{Dea_ea:2009}.

\section{\label{ionic_liquid}Ionic liquid lattice capacitor}

In the models above the ions have been chosen to be point-like. Here we address the question of changes wrought by their finite size. In this case the system consists of a 1D lattice of $M$ sites with spacing $a$, with the $i-$th site, $0 \le i < M$,  occupied by ion with charge $qS_i$ with $S_i \in [-q,0,q]$, see Fig. \ref{scheme} (bottom). Within this model the finite ion size is $\sim a$, which is crucial to the phenomena observed in experiments on confined ionic liquids. 

The configuration described is one of the 1D {\em ionic liquid capacitor}. The external  fields are imposed either by fixing the charges of the boundaries at $i = -1$ and $i = M$ to be $\pm qQ$, respectively, or by  imposing a fixed voltage/potential difference, $\Delta v$, across the capacitor. More details can be found in \cite{Dem_ea:2012}. 

The electrostatic Hamiltonian in this case is expressed through a spin-like variable $S_i = 0,\pm 1$
\[
\beta{\cal H} = -\frac{\gamma}{4}\sum_{i.j=0}^{M-1}|i-j|S_iS_j~~~~~~\gamma = \frac{\beta q^2a}{\varepsilon}.
\]
After a Hubbard-Stratonovich transformation this yields the action
\[
S(\phi) = \sum_{j=0}^{M-2} \frac{(\phi_{j+1}-\phi_j)^2}{2\gamma} - \sum_{j=0}^{M-1}\ln[1+2\mu\,\cos(\phi)] + iQ(\phi_{-1}-\phi_M).
\]
The system includes boundary charges $\pm qQ$ at sites $-1, M$.  The electrostatic potential is defined as $V = -i\phi/\beta q$. In limit  $a \to 0,~q/a$ fixed, the  MF equations obtained from the saddle-point of the above field action reduce to those of Kornyshev \cite{Kor:2007} and Borukhov et al. \cite{Bor_ea:2000}.


For non-zero $a$ the action is not positive definite for $\mu \ge 0.5$ and so we seem to have a sign problem and certainly cannot use the Schr\"odinger approach {\em a priori}. Nevertheless, in the case of 1D the partition function can be computed exactly by using the transfer matrix approach, with the Fourier method described earlier. This can be seen as follows: write $y_i = \phi_i$ and define
\[
p^{1/2}(y,y') = \frac{1}{\sqrt{\pi \gamma}}e^{-(y-y')^2/\gamma},
\]
with
\begin{eqnarray}
K(y,y')&=&\int dz\;p^{1/2}(y,z)[1+\mu\,\cos(z)]p^{1/2}(z,y')
\end{eqnarray}
with $Kf(y) = \int_{-\infty}^\infty dy'\;K(y,y')f(y')$. The free energy for the fixed $Q$ ensemble, $\Omega_Q$, then follows as 
\[
e^{-\beta\Omega_Q} = \int_{-\pi}^\pi dx\; e^{iQx}\left[p^{1/2}K^Mp^{1/2}\right]e^{-iQx}~\equiv~\langle \psi_Q|K^M|\psi_Q\rangle
\]
with 
\[
\langle \psi_Q|y \rangle =  e^{iQy},~~~~\langle y|K| y'\rangle = K(y,y')
\]
 The conjugate free energy for the fixed $\Delta v$ ensemble, $\Omega_{\Delta v}$, then follows from a Legendre transform, 
\[
e^{\textstyle -\beta\Omega_{\Delta v}} = \int dQ\;e^{-\Delta v Q - \beta \Omega_Q},
\]
while the capacitance $C_{\Delta v}$ is obtained from the first derivative of $\partial \langle Q \rangle_{\Delta v}$ w.r.t. $\Delta v$.
and can thus be calculated directly from the partition function.

\subsection{Results}
The transfer matrix and Fourier approach can be formulated in order to evaluate the free energy explicitly. Details of this procedure can be found in Ref. \cite{Dem_ea:2012}.
Enthalpy ${\cal G}_M = \Omega_M + MP_\mathrm{bulk}$, the disjoining pressure $P = {\cal G}_M - {\cal G}_{M+1}$, and the capacitance $C_{\Delta v}$, can all be calculated as a function of $\mu, Q, \Delta v$. 

\begin{figure*}[t!]\begin{center}
\begin{minipage}[b]{0.45\textwidth}\begin{center}
\includegraphics[width=\textwidth]{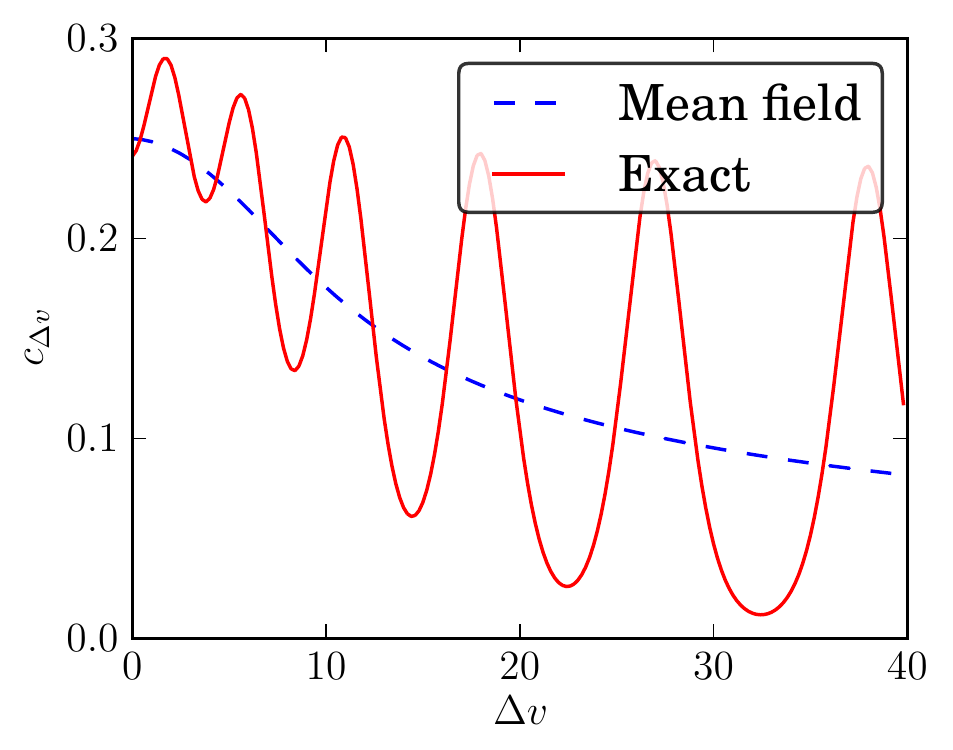} 
\end{center}\end{minipage}
\begin{minipage}[b]{0.45\textwidth}\begin{center}
\includegraphics[width=\textwidth]{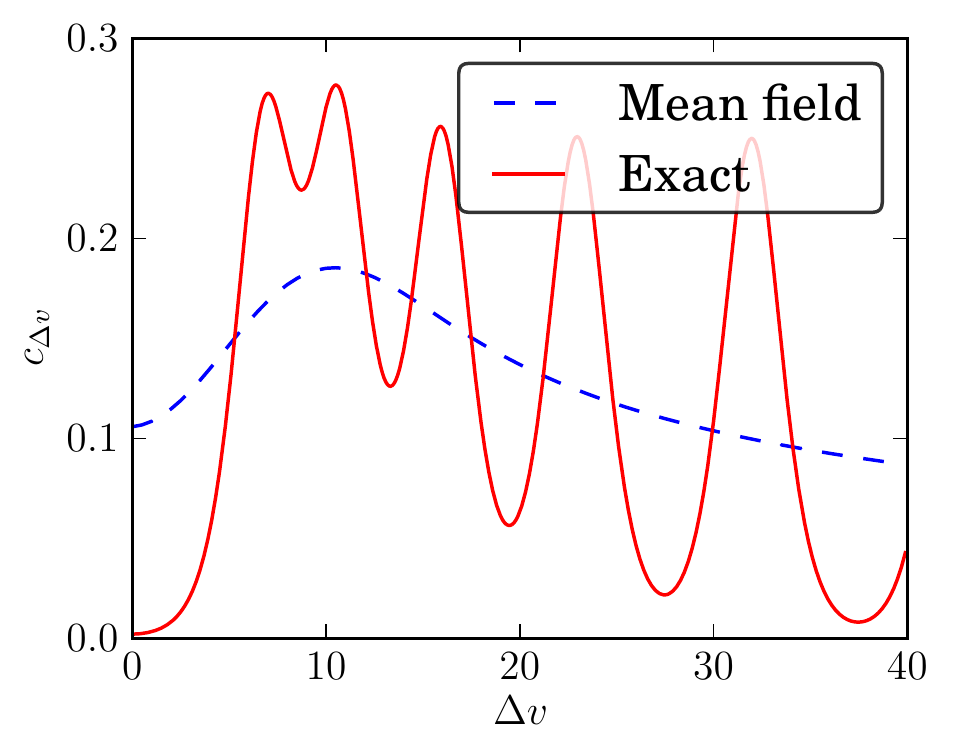} 
\end{center}\end{minipage}
\caption{(Left) Bell shaped $C_{\Delta v}$ as a function of $\Delta v$ for large $\mu= 0.5$. (Right) Camel shaped $C_{\Delta v}$ as a function of $\Delta v$ for small $\mu= 0.03$. In both cases, we have $\gamma = 1$. The dashed line is the MF result \cite{Dem_ea:2012}.}
\label{fig:C_bell}
\end{center}\end{figure*}

We show explicitly only the capacitance results, $C_{\Delta v}$, as a function of $\Delta v$ in Fig. \ref{fig:C_bell}, both for large $\mu$ and small $\mu$. For large $\mu$ the curve shows shows the typical ``bell'' shape in contrast to the curve for smaller $\mu$, which shows the non-monotonic ``camel'' shape  and so $C_{\Delta v}$ has a minimum at the point of zero charge confirming the Fermi MF results of Kornyshev \cite{Kor:2007}. 

For smaller $\gamma$ (increasing $T$) the periodic non-monotonicity both for large $\mu$ and small $\mu$ disappears and the solution approaches the Fermi MF result of Kornyshev \cite{Kor:2007}. It is interesting that the exact solution dances around the Kornyshev solution with an ever increasing amplitude but the system nevertheless always remains thermodynamically stable, as can be straightforwardly ascertained.

\section{\label{conclusions}Lessons}
We have demonstrated that in 1D one can use the Schr\"odinger approach for continuum models of Coulomb fluids,
but that for discrete models a more general approach is needed which exploits the transfer matrix and the periodicity
of the field to use Fourier methods.  We tested the PB/MF and the strong coupling limiting expressions  and demonstrated that they need correcting although the exact  analytic result clearly supports the two limiting analyses. We also confirmed that the MF theory does not capture the important effects which are due to correlations, either the attractive intersurface forces in the case of a counterion-only system or non-monotonic periodic variation of the capacitance in the confined ionic liquid case. 


\bibliography{cecam} \bibliographystyle{unsrt}

\begin{thebibliography}{10}

\bibitem{netz:2005}
H.~Boroudjerdi, Y.-W. Kim, A.~Naji, R.~R. Netz, X.~Schlagberger, and A.~Serr.
\newblock Statics and dynamics of strongly charged soft matter.
\newblock {\em Phys. Rep.}, {\bf 416}:129, 2005.

\bibitem{EdwLen:1962}
S.~F. Edwards and J.~Lenard.
\newblock {Exact Statistical Mechanics of a 1 Dimensional System with Coulomb
  Forces. 2. Method of Functional Integrals}.
\newblock {\em J. Math. Phys.}, {\bf 3}:778, 1962.

\bibitem{PodZek:1988}
R.~Podgornik and B.~\v{Z}ek\v{s}.
\newblock {Inhomogeneous Coulomb Fluid -- A Functional Integral Approach}.
\newblock {\em J. Chem. Soc. -- Farad. Trans.}, {\bf 84}:611, 1988.

\bibitem{Dea_ea:1998}
D.~S. Dean, R.~R. Horgan, and D.~Sentenac.
\newblock {Boundary Effects in the One Dimensional Coulomb Gas}.
\newblock {\em J. Stat. Phys.}, {\bf 90}:899, 1998.

\bibitem{NetOrl:1999}
R.~R. Netz and H.~Orland.
\newblock {Field Theory for Charged Fluids and Colloids}.
\newblock {\em Europhys. Lett.}, {\bf 45}:726, 1999.

\bibitem{DeaHor:2005}
D.~S. Dean and R.~R. Horgan.
\newblock The field theory of symmetrical layered electrolytic systems and the
  thermal casimir effect.
\newblock {\em J. Phys. C}, {\bf 17}:3473, 2005.

\bibitem{Dea_ea:2009}
D.~S. Dean, A.~Naji, and R.~Podgornik.
\newblock One-dimensional counterion gas between charged surfaces: Exact
  results compared with weak- and strong-coupling analysis.
\newblock {\em J. Chem. Phys.}, {\bf 130}:094504, 2009.

\bibitem{Dem_ea:2012}
V.~Demery, D.~S. Dean, T.~C. Hammant, R.~R. Horgan, and R.~Podgornik.
\newblock The one-dimensional coulomb lattice fluid capacitor.
\newblock {\em J. Chem. Phys.}, {\bf 137}:064901, 2012.

\bibitem{Kor:2007}
A.~A. Kornyshev.
\newblock {Double-Layer in Ionic Liquids: Paradigm Change?}
\newblock {\em J. Phys. Chem. B}, {\bf 111}:5545, 2007.

\bibitem{Bor_ea:2000}
I.~Borukhov, D.~Andelman, and H.~Orland.
\newblock {Adsorption of large ions from an electrolyte solution: a modified
  Poisson-Boltzmann equation}.
\newblock {\em Electrochem. Acta}, {\bf 46}:221, 2000.

\end{thebibliography}

\end{document}